\documentclass[aps,pre,11pt,onecolumn,nofootinbib]{revtex4-2}

\usepackage{amsmath,amssymb,bm}
\usepackage{mathptmx}
\usepackage[colorlinks=true,linkcolor=blue,citecolor=blue,urlcolor=blue]{hyperref}
\usepackage{graphicx}

\def\la{\langle}
\def\ra{\rangle}
\begin{document}

\title{Heat capacity as a marker for\\ shape and jamming transitions in active systems}
\author{Ion Santra 
} 
\affiliation{Department of Physics and Astronomy, KU Leuven, Belgium}

\begin{abstract}
Persistence influences the stationary states of active particles, producing boundary accumulation at the single-particle level, and clustering or jamming in interacting systems. These features disappear as the persistence decreases and the system approaches a more passive-like stationary state. We show that these transitions have a distinct calorimetric signature. Using a lattice run-and-tumble dynamics, consistent with local detailed balance, we compute the nonequilibrium heat capacity from the excess heat released
following a small temperature perturbation. For a single particle confined between reflecting boundaries, the heat capacity develops a maximum in the persistence regime corresponding to shape transition. Adding an exclusion interaction to the active particles 
on a periodic lattice, the reorganization
of jammed clusters produces a corresponding peak in the thermal response. 
We also discuss the impact of the time-symmetric part of the transition rates, and show the possibility of seeing the same signatures of heat response in experiments by AC calorimetry. Our results show that nonequilibrium heat capacities can serve as calorimetric probes of nonequilibrium phase transitions.
\end{abstract}
\maketitle
\section{Introduction}

Active particles have emerged as paradigmatic examples of systems driven far from equilibrium. They include self-propelled objects over a broad range of
scales, from bacteria, motile cells, and synthetic microswimmers to vibrated grains and robotic agents~\cite{romanczuk2012,bechinger2016active}. The self-propulsion, which is persistent, is maintained by a continuous consumption of energy at the level of the individual particle. Consequently, their dynamics cannot be described in terms of thermal fluctuations and conservative forces alone, and violate detailed balance~\cite{fodor2016far,activeparticle1,demaerel2018active}. Already at the single-particle level, persistence produces rich athermal fluctuations~\cite{Basu_Majumdar_Rosso_Schehr_2019,santra2020run,santra2021active,santra2022universal}, for example, in a confining potential, active
particles may develop strongly non-Boltzmann density profiles and, in some
cases, enhanced occupation of regions that would be energetically unfavorable
at equilibrium~\cite{rtp_distribution,basu2018active,santra2021direction}. In the presence of hard boundaries, they typically accumulate
near the boundaries~\cite{bechinger2016active,li2009accumulation,elgeti2015run}. When the persistence time decreases there is a shape transition of the distribution from \emph{active}-like boundary crowding to passive behavior. Persistence also has very interesting consequences in interacting systems. Active particles subject to repulsive interactions undergo motility-induced phase separation~\cite{fily2012athermal,redner2013structure,cates2015motility}. In one dimension, where particles cannot pass each other, the same competition between persistent propulsion and exclusion produces long-lived jammed
clusters~\cite{soto2014run,slowman2016jamming,dandekar2020hard,mukherjee2023nonexistence}. A decrease in persistence  destabilizes these structures, and drives the system towards a more homogeneous fluid-like state.

These observations raise a natural thermodynamic question: does the reorganization of the stationary state leave a measurable heat signature? 
Though the thermodynamics of active particles has received considerable attention in the last decade~\cite{ganguly2013stochastic,speck2016stochastic,marconi2017heat,padmanabha2023fluctuations,pruessner2025field,paoluzzi2024entropy}, the focus has been mainly on how activity affects entropy production, fluctuation relations, etc. Here, instead, we are interested in whether thermodynamic response can reveal the persistence-driven transitions in active systems.
Since an active system continuously dissipates energy even in its stationary state, the total heat released following a temperature change contains a housekeeping contribution that grows with time. The relevant calorimetric
quantity is therefore the excess heat associated with relaxation from one nonequilibrium stationary state to another~\cite{oono1998steady,hatano2001steady}. Its differential response to a small temperature perturbation defines a nonequilibrium heat capacity~\cite{pevsek2012model,epl,calo}. This has been studied in a variety of driven systems, including few-state and periodically driven models, active particles, and heat-transport setups~\cite{epl,cejp,calo,elena1,elena2,gautama}. Recent work has further clarified the origin of negative nonequilibrium heat capacities~\cite{bogers2025negative}, while calorimetry has also been proposed as a means of probing activity in active-particle and minimal biophysical models~\cite{dolaiactivecalo,khodabandehlou2026bringing}.  In this paper, we focus on a complementary question, whether well-known persistence driven transitions in active systems, namely, shape transition for a single confined particle, or the jamming transitions for interacting active particles, leave identifiable calorimetric signatures.


We introduce a
one-dimensional lattice model of run-and-tumble particles whose hopping rates satisfy local detailed balance. We first consider a single particle confined
between reflecting boundaries. At small tumbling rates, the stationary density is strongly concentrated near the walls, whereas rapid tumbling produces an
approximately uniform profile. The heat capacity develops a pronounced maximum in the same range of tumbling rates in which the boundary accumulation is lost.  Secondly, we consider interacting run-and-tumble particles with hard-core exclusion on a periodic lattice. At large persistence, the particles form extended jammed structures, while rapid tumbling leads to a more homogeneous stationary state. We show that, the reorganization is accompanied by a
nonmonotonic excess-heat response. At sufficiently high density, even the initial enhancement of clustering leaves a separate calorimetric feature.

Our central result is that structural transitions in active systems are accompanied by pronounced features in the nonequilibrium heat capacity. We also show that the detailed form of the response is not universal, as is generally expected out of equilibrium~\cite{epl,cejp,bogers2025negative}. Changing the temperature-dependent kinetic prefactor can modify its magnitude and enhance peaks, dips, or negative values. Nevertheless, the strongest calorimetric variations remain tied to the regime in which the stationary structure changes most rapidly. The heat capacity therefore provides a thermodynamic probe of persistence driven active-passive transition.

The remainder of the paper is organized as follows. We first introduce the locally detailed-balanced lattice dynamics in Sec.~\ref{s:model}. We then study the stationary
distribution and excess heat of a single particle between reflecting
boundaries in Sec.~\ref{s:single}.  We move to interacting particles on a
periodic lattice in Sec.~\ref{s:interact} to illustrate the persistence induced jamming to unjamming transition, and obtain the corresponding calorimetric response. We examine how temperature-dependent kinetic prefactors affect
the structural and calorimetric signatures in Sec.~\ref{sec:kineticfactor}. In Sec.~\ref{s:accal}, we discuss how to access these heat responses experimentally. Finally, we conclude in Sec.~\ref{s:concl}




\section{Non-interacting run-and-tumble particle on a lattice}\label{s:model}
\begin{figure}
    \centering
    \includegraphics[width=0.48\linewidth]
    {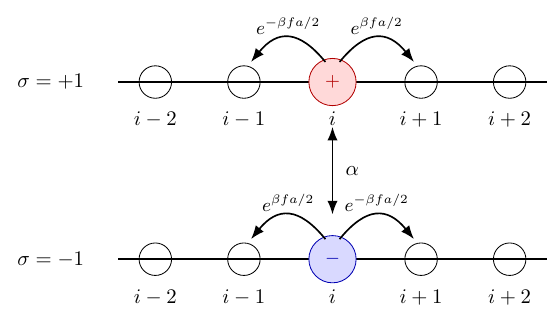}
    \includegraphics[width=0.35\linewidth]{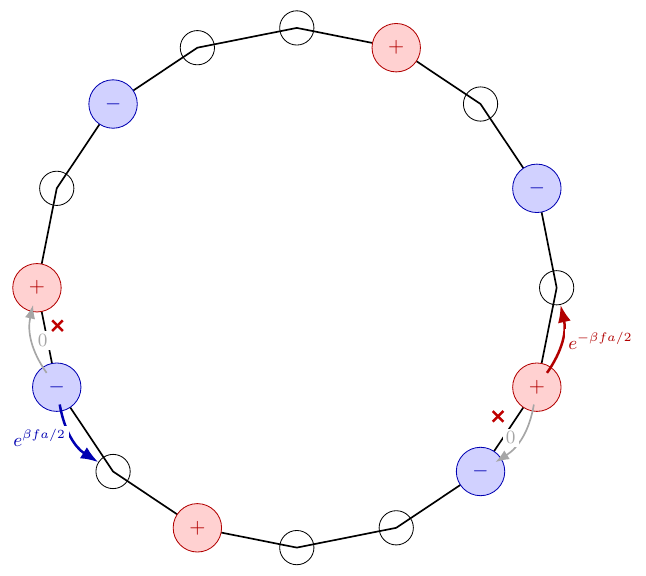}
    \caption{Schematic representation of the single-particle and interacting
lattice models, depicting the hopping and tumbling rates.}
    \label{fig:placeholder}
\end{figure}

We begin with a single run-and-tumble particle moving on a one-dimensional lattice in continuous time. The particles have an internal orientation $\sigma$ which biases its direction of movement on the lattice. The random variable $\sigma$ is two-state Markov jump process with states $\pm 1$, and the rate of switching between each other (which is the tumbling in this case) being the same constant $\alpha$. Note that, since there is no bias in the $\sigma$-space, the states are energetically degenerate and there is no energy exchange associated with it. The spatial jumps $i\to i+\delta$ [$\delta=\pm1$] occur with transition
rates $k(i,\sigma\to i+\delta,\sigma)$. We assume that these rates satisfy the
local detailed balance condition~\cite{maes2021local}\footnote{A different thermodynamically consistent realization of run-and-tumble dynamics was proposed in Ref.~\cite{khodabandehlou2024local}, where the positional and internal degrees of freedom are coupled to baths at different temperatures. That two-temperature model also displays a persistence-dependent shape transition. Here, instead, the tumbling rate is constant and local detailed balance is imposed on the spatial hopping rates.},
\begin{equation}
\frac{k(i,\sigma\to i+\delta,\sigma)}{k(i+\delta,\sigma\to i,\sigma)}
=
\exp\left[
\beta
\left(
f a\sigma\delta-\Delta U_i^\delta
\right)
\right],
\label{eq:single-rate}
\end{equation}
 In the above equation, $a$ is the lattice constant, and $ \Delta U_i^\delta=U_{i+\delta}-U_i$. While Eq.~\eqref{eq:single-rate} fixes only the ratio of the forward and backward
rates, they could be realized in many ways. We first choose the symmetric rates
\begin{equation}
    k(i,\sigma\to i+\delta,\sigma)
=\nu
\exp\left[
\frac{\beta}{2}
\left(
f a\sigma\delta-\Delta U_i^\delta
\right)
\right],
\label{eq:single-rate2}
\end{equation}
with $\nu$ being independent of temperature. The effect of allowing this prefactor to depend on
temperature, $\nu=\nu(T)$, is discussed later in
Sec.~\ref{sec:kineticfactor}.
It is worth pointing out that the difference with the usual model of $1$-dimensional continuous space RTPs lies in the fact that, here, even in the force free case ($U_i=0,\,\forall i\in[-L,L]$), an RTP with a positive orientation can make a leftwards jump, though with a very small probablity (similar considerations also appeared in \cite{angelani2023one,padmanabha2023fluctuations}). Only in the limit $\beta\to\infty$ does the dynamics approach that of the
usual lattice RTP, where the particle is allowed to move only along the direction of $\sigma$. Finally, to conclude the model setup, 
 we note that the heat released to the environment in a spatial jump is 
\begin{equation}
q(i,\sigma\to i+\delta,\sigma)
=
f a\sigma\delta-\Delta U_i^\delta .
\label{eq:single-jump-heat}
\end{equation}

Note that the continuum limit of this lattice RTP is obtained by taking $a\to0$ and
$\nu\to\infty$, while keeping $D=\nu a^2$ fixed. 
Writing $x=ia$ and $U_i=U(x)$, we get
\begin{align}
\dot x(t)
&=
\beta D
\left[
f\sigma(t)-U'(x)
\right]
+
\sqrt{2D}\,\xi(t),\label{eq:contiunuum}
\end{align}
with 
$\sigma(t)=\pm 1$ flipping as before. Thus, in contrast to the usual thermal RTP where the active speed is taken to be independent of temperature, here the propulsion velocity is $\beta Df$. 

\section{Single particle on a finite line: shape transitions}\label{s:single}
In this section, we consider a single RTP, with the dynamics described in the previous section on a finite lattice interval $[-L,L]$. 
\subsection{Stationary distribution}
The RTP dynamics described above is non-Markovian in the position alone, but Markovian when written in the extended state space $(i,\sigma)$. It is useful to formulate the dynamics in terms of its forward generator. We denote by
$\mathcal{L}^\dagger$ the forward generator acting on the probability distribution 
$\mathcal P(t)=(p_i(t),m_i(t))$, where $p_i(t)$ and $m_i(t)$ are the probabilities of finding the particle
at site $i$ with orientations $+1$ and $-1$, respectively, at time $t$. The master equation 
is given by, $\partial_t \mathcal P=\mathcal{L}^\dagger \mathcal P$.
In component form, 
\begin{subequations}
\begin{align}
\dot p_i(t) &=
R_{i-1}^{+}p_{i-1}(t)
+
L_{i+1}^{+}p_{i+1}(t)
-
(R_i^+ + L_i^+)p_i(t)
+
\alpha(m_i(t)-p_i(t)),
\label{eq:p-master}
\\
\dot m_i(t) &=
R_{i-1}^{-}m_{i-1}(t)
+
L_{i+1}^{-}m_{i+1}(t)
-
(R_i^- + L_i^-)m_i(t)
+
\alpha(p_i(t)-m_i(t)),
\end{align}\label{eq:m-master}
\end{subequations}
where, the hopping rates are given by (with $\nu=1$),
\begin{align}
    R^+_{i}=e^{\frac{\beta}{2}(fa-\Delta U^+_i)},\qquad L^+_{i}=e^{\frac{\beta}{2}(-fa-\Delta U^-_i)},\quad R^-_{i}=e^{\frac{\beta}{2}(-fa-\Delta U^+_i)},\qquad L^-_{i}=e^{\frac{\beta}{2}(fa-\Delta U^-_i)}.
\end{align}

 In the stationary state, $\dot p_i(t)=\dot m_i(t)=0$,  giving 
 the stationary distribution 
 $P_T(i,\sigma)$, where $p_i=P_T(i,+1)$, $m_i=P_T(i,-1)$, and $\rho_i=p_i+m_i$ (where the subscript  $T$ denotes the temperature set by the local detailed balance rates Eq.~\eqref{eq:single-rate}). 
 We need to solve the above Eqs.~\eqref{eq:m-master} with the boundary rates $R^{\sigma}_L=L^{\sigma}_{-L}=0$, accounting for the hard walls. Moreover,  terms involving sites outside the interval $[-L,L]$ are omitted as and when they appear
 for the boundary probabilities. Adding the two stationary  equations,  
 we get an equation for the marginal position distribution $\rho_i$ upon setting $\dot\rho_i=0$:
\begin{align}
    0=-J_i+J_{i-1},\quad\text{with}~~  J_i=j_i^+ + j_i^-
    \label{eq:dotrhoeq0}
\end{align}
where $J_i$ denotes the total currents and $j_i^\sigma$ are the oriented currents across the $(i,i+1)$th bond given by,
\begin{align}
j_i^+
&=
R_i^+ p_i-L_{i+1}^+p_{i+1},\label{jiplus}
\\
j_i^-
&=
R_i^- m_i-L_{i+1}^-m_{i+1}.\label{jiminus}
\end{align}
Using Eq.~\eqref{eq:dotrhoeq0} and the fact that there are hard boundaries, we conclude that $J_i=0$ implying  $j_i^+ =- j_i^-$.  Furthermore, from Eq.~\eqref{eq:p-master} in the stationary state, we get,
\begin{align}
    j_i^+=j_{i-1}^+ + \alpha(m_i-p_i).
\end{align}
Combining the above with Eq.~\eqref{jiplus} and \eqref{jiminus}, we can recast it as
\begin{align}
    X_{i+1}=\mathcal T_i X_i
\end{align}
where $X_i^T=(p_i~m_i~j^+_{i-1})$, and $\mathcal T_i=\begin{pmatrix}
    \dfrac{R_i^+ + \alpha}{L_{i+1}^+}
&
-\dfrac{\alpha}{L_{i+1}^+}
&
-\dfrac{1}{L_{i+1}^+}
\\[1.2em]
-\dfrac{\alpha}{L_{i+1}^-}
&
\dfrac{R_i^- + \alpha}{L_{i+1}^-}
&
\dfrac{1}{L_{i+1}^-}
\\[1.2em]
-\alpha
&
\alpha
&
1
\end{pmatrix}$. 
The solution is then,
\begin{equation}
    X_i=\mathcal T_{i-1}\mathcal T_{i-2}\cdots \mathcal T_{-L}X_{-L}
\end{equation}
with the boundary condition, $X_{-L}^T=(p_{-L}~m_{-L}~0)$. The other necessary conditions include the boundary condition $j^+_L=0$ and the normalization $\sum_{i=-L}^L(p_i+m_i)=1$.

\begin{figure}
    \centering
    \includegraphics[width=0.45\linewidth]{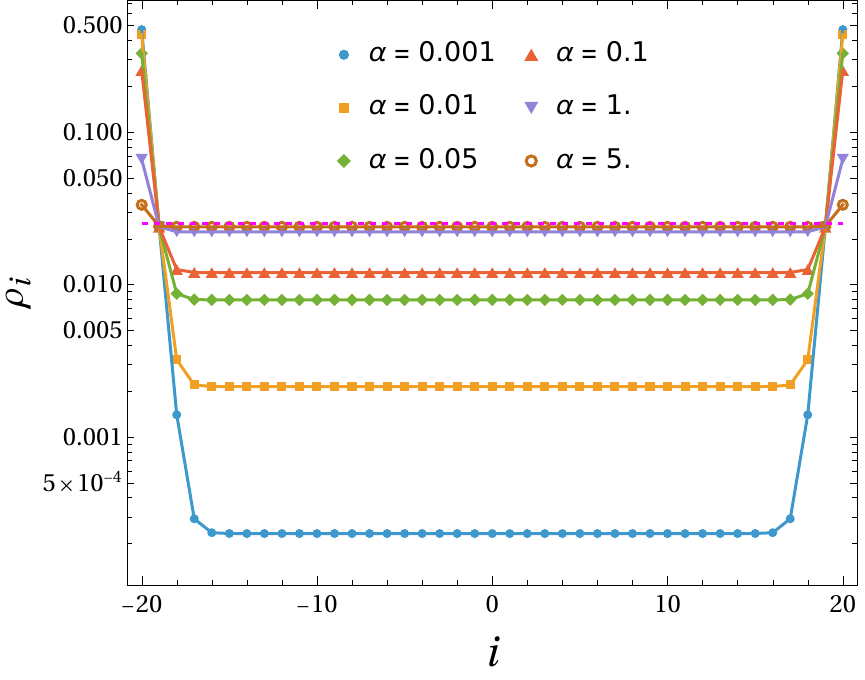}
    \includegraphics[width=0.45\linewidth]{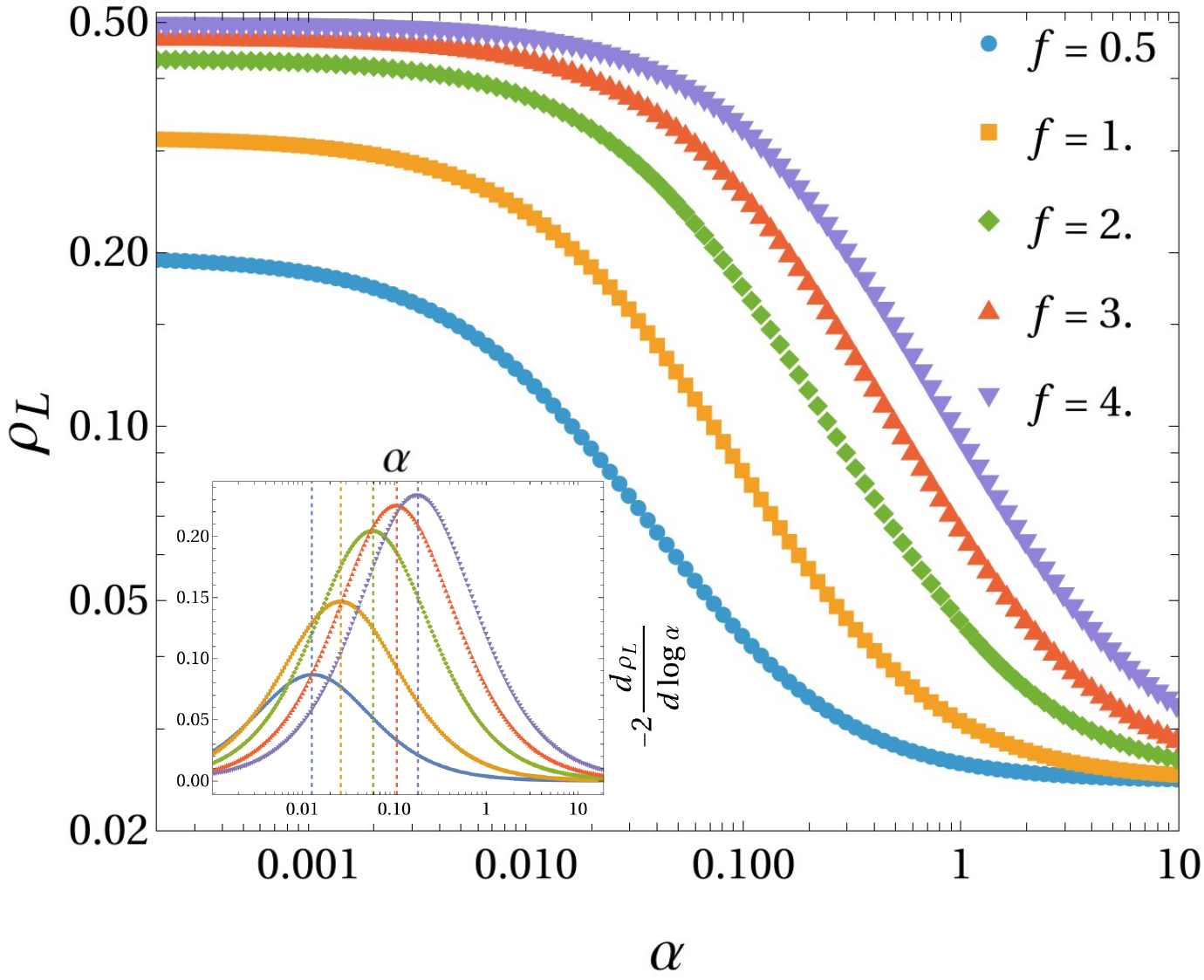}
    \caption{Left: Stationary density $\rho_i$ for different tumbling
rates $\alpha$ at $f=2$. Right: Boundary density $\rho_L$ as a function of $\alpha$
for different $f$; the inset shows
$-d\rho_L/d\log\alpha$, with the vertical lines given by Eq.~\eqref{alphastar}. For both plots, we take $L=20$, $T=1$, $a=1$, and $\nu=1$.}
    \label{f:singleparticless}
\end{figure}

 In the remainder of this section, we set $U_i=0$ .Then $R_i^+=L_i^-=e^{\beta fa/2}$, and $R_i^-=L_i^+=e^{-\beta fa/2}$, and we have,
\begin{equation}
    X_i=\mathcal T^{i+L}X_{-L}.
\end{equation}
The density at the $i$th site is given by 
\begin{equation}
    \rho_i=(1~1~0)\mathcal T^{i+L}\begin{pmatrix}
        p_{-L}\\
        m_{-L}\\
        0
    \end{pmatrix}
\end{equation}
Moreover, at the right boundary $j_L^+=0$, implying $(-\alpha ~\alpha~1)X_L=0$. This leads to
\begin{align}
    \frac{m_{-L}}{p_{-L}}=-\frac{b^T \mathcal T^{2L}e_1}{b^T \mathcal T^{2L}e_2}=\Lambda 
\end{align}
where $b^T=(-\alpha~ \alpha ~ 1)$ $e_1^T=(1~0~0)$ and $e_2^T=(0~1~0)$. 
Using normalization, we get,
\begin{align}
    \rho_i=Z^{-1} (1~1~0)\mathcal T^{i+L}(e_1+\Lambda e_2),\quad \text{with }Z=\sum_{i=-L}^{L} (1~1~0)\mathcal T^{i+L}(e_1+\Lambda e_2).
\end{align}
The stationary profile plotted for different values of $\alpha$ are shown in the left panel of Fig.~\ref{f:singleparticless}. We see that the stationary profile shows a boundary accumulation for small values of $\alpha$, and tends to the uniform value $1/(2L+1)$, as $\alpha$ becomes larger.

The loss of boundary accumulation can be quantified by monitoring the occupation of a boundary site, shown in the right panel of
Fig.~\ref{f:singleparticless}. At small tumbling rates, the particle typically reaches the wall before losing its orientation, and stays there until a tumble event, the boundary
density $\rho_L$ is therefore enhanced. 
As $\alpha$ increases, the particle typically undergoes much more tumbling events, re-orienteering it before covering the full wall-to-wall distance. As a result the boundary excess goes down gradually, saturating to the uniform value $\rho_L=1/(2L+1)$.

Since this change takes place over a broad range of tumbling rates, we measure the variation with respect to $\log\alpha$ via $-\frac{d\rho_L}{d\log\alpha}$, which shows a distinct maxima at some 
$\alpha=\alpha^*$. The position of this maximum can be estimated from a persistence-length argument. For a particle with a particular orientation, say $\sigma=+1$, the effective drift speed is
$v_{\text{drift}}=2\nu\sinh(\beta fa/2)$, measured in lattice sites per unit time.
The typical run length before a tumble is then $v_{\text{drift}}/\alpha$. We expect the boundary accumulation to be high as long as the total lattice length $2L$ is smaller than this persistence length. This gives a crossover value $\alpha^*$
\begin{equation}
\alpha^*
\simeq
\frac{\nu\sinh\left(\frac{\beta f a}{2}\right)}{L}.\label{alphastar}
\end{equation}
This prediction agrees very well with the maximum of
$-d\rho_L/d\log\alpha$ shown in the inset of the right panel of
Fig.~\ref{f:singleparticless}.


\subsection{Excess heat}
We now ask whether this shape transition has a calorimetric signature. For this purpose we compute the nonequilibrium heat capacity from the excess heat, using the formalism of nonequilibrium calorimetry, introduced by Maes and Neto{\v{c}}n{\'y} in~\cite{calo}. Since the active force continuously dissipates heat in the NESS, the heat produced during a temperature change contains an extensive housekeeping contribution that is needed to maintain the NESS. The nonequilibrium heat capacity is instead defined from the excess heat associated with the relaxation between the NESSs.

Let $X=(i,\sigma)$ denote the state of the particle. The mean instantaneous heat flux to the environment, conditioned on the state $X$ is
\begin{align}
    \dot Q_T(X)=\sum_Y k_T(X,Y) q(X,Y)
    \label{eq:dotq1}
\end{align}
where $q(X,Y)$ is given in Eq.~\eqref{eq:single-jump-heat}. We define the centered heat flux 
\begin{equation}
    h_T(X)=\dot Q_T(X)-\la\dot Q_T\ra\quad\text{with  }~ \la\dot Q_T\ra=\displaystyle \sum_X P_T(X)\dot Q_T(X) 
    \label{eq:centered-heat-flux}
\end{equation}
 which has a zero stationary average. 
 For $U_i=0$, the centered heat flux can be written in an insightful form . Let $B_\text{out/in}(X)$ indicate that the particle has an outward (inward) orientation at the boundary, and further $B=B_\text{out}+B_\text{in}$, $M=B_\text{out}-B_\text{in}$. Then the centered heat flux is given by [see App.~\ref{app:1}],
 \begin{align}
     h_T(X)=-\frac{fa}{2}\left( (r-l)\delta B(X)+(r+l)\delta M(X) \right), \quad \delta O=O-\la O\ra.
     \label{inst:excess-single}
 \end{align} 
 Thus, though the active particle dissipates heat throughout the lattice, the state dependence of the excess instantaneous heat comes from the boundary layers. 
 
 Starting from a state $X$, the mean excess heat released is given by the quasipotential,
\begin{equation}
    V_T(X)=\int_0^\infty dt \left[ \la\dot Q_T(X_t)\ra_X-\la\dot Q_T\ra  \right]
\end{equation}
Operating the backward generator $\mathcal{L}$ on both sides, we arrive at the Poisson equation for the quasipotential~\cite{khodabandehlou2024poisson},
\begin{equation}
    \mathcal{L}V_T(X)=-h_T(X),
\end{equation}
 with the contraint $\la V_T(X)\ra=\sum_XP_T(X)V_T(X)=0.$ For an infinitesimal temperature change $T\to T+dT$, the corresponding excess heat released to the environment is
\begin{align}
    dQ=\sum_X \left[P_T(X)-P_{T+dT}(X)\right]V_T(X)=-dT\sum_X \frac{\partial P_T(X)}{\partial T}V_T(X)
\end{align}
The nonequilibrium heat capacity which is the excess heat released per $dT$ is the negative differential of the above quantity with respect to temperature,
\begin{equation}
    C=-\frac{dQ}{dT}=\sum_X \frac{\partial P_T(X)}{\partial T}V_T(X)=-\left\la\frac{\partial V}{\partial T}\right\ra 
    \label{eq:spheatdef}
\end{equation}

We plot the heat capacity as a function of $\alpha$ for different values of the active force magnitude $f$ (the details of the numerics are given in App.~\ref{app:numerics}). We find that for very small $\alpha$, the particle is strongly persistent and the boundary accumulation changes very slowly (see Fig.~\ref{f:singleparticless}), so a small temperature change does not reorganize the stationary state very efficiently. 
On the other hand, for very large $\alpha$, there is rapid tumbling and the stationary state becomes close to uniform, and changing temperature does not affect the NESS much. 
The largest response occurs between these two limits, when the persistence length is comparable to the system size. This is evident from the maxima observed at intermediate values of $\alpha$. In fact, we see that the maximum occurs close to the same value where the boundary occupation changes the most rapidly, namely at $2\alpha^*/\pi$ \footnote{Note that the factor $2/\pi$ is inferred from the numerical results, that provides an excellent estimate of the location of the maximum across different parameter regimes.}.
In that regime, changing the temperature changes the effective drift strength and hence strongly modifies the boundary excess. 
The relaxation following such a perturbation therefore carries a large excess heat. 
\begin{figure}
    \centering
    \includegraphics[width=0.65\linewidth]{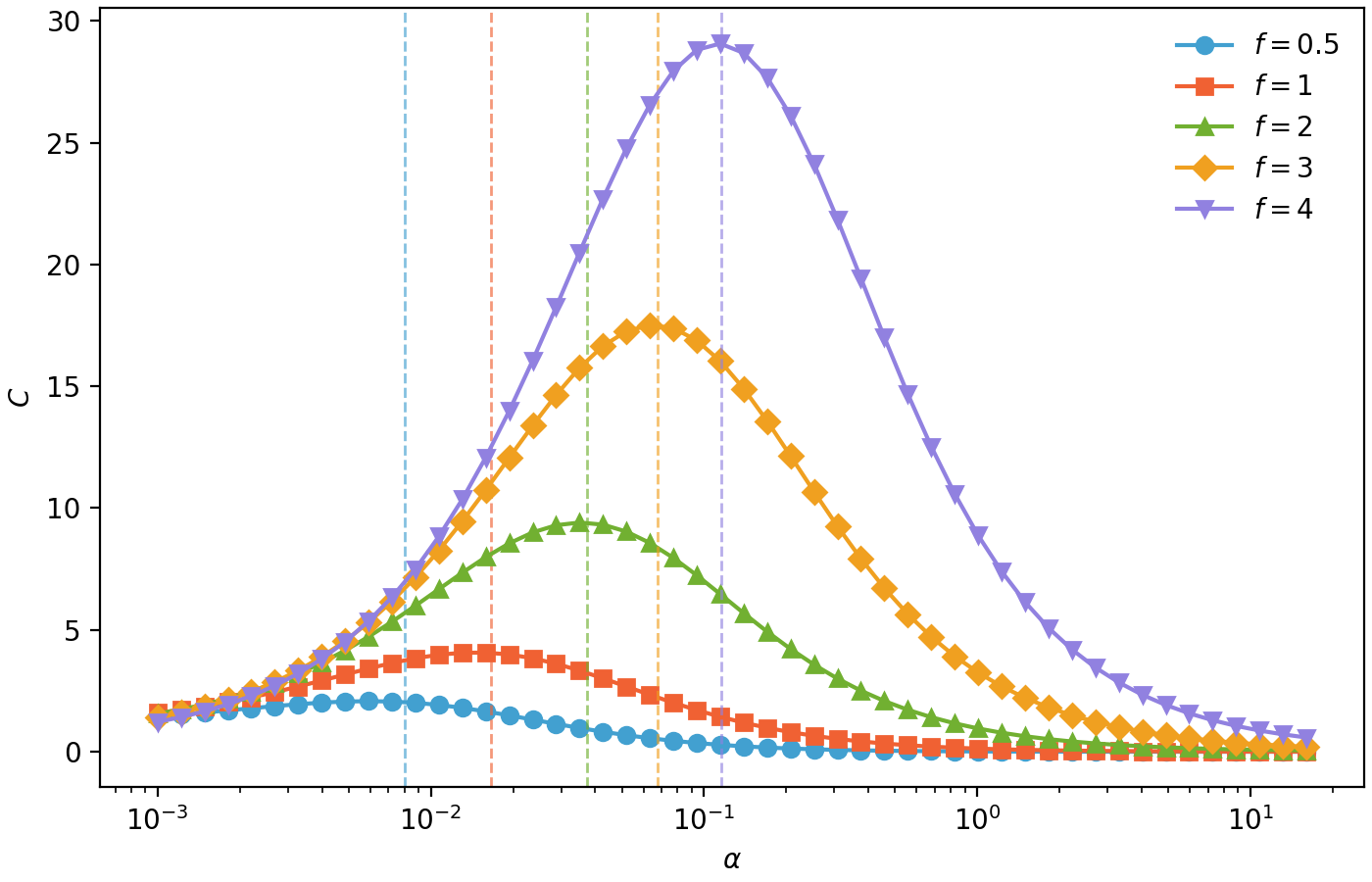}
    \caption{Nonequilibrium heat capacity $C$ as a function of $\alpha$ for
different active forces $f$.
The dashed lines indicate $2\alpha^*/\pi$.
Here $L=20$, $T=1$, $a=1$, and $\nu=1$.}
    \label{fig:spheat-single}
\end{figure}


\section{Interacting RTPs on a periodic lattice}\label{s:interact}
\subsection{Model and stationary states}
We 
turn to the interacting case, and 
we remove the confining boundaries by placing the particles on a periodic lattice. This eliminates boundary accumulation as the relevant structural effect. Instead, the persistence of the active motion now competes with the hard-core interaction between particles, and this gives rise to jamming.

We consider $N_p$ run-and-tumble particles on a ring of $L$ sites. It is convenient to describe a configuration as $\Omega=(\tau_1,\dotsc,\tau_L)$ with 
 $\tau_i=0$ if site $i$ is empty, and $\tau_i=\pm 1$ if occupied by a $\sigma=\pm1$ particle. Then, $n_i=\tau_i^2=\{0,1\}$ denotes the occupation of the $i$th site. 
The allowed transition rates are defined as follows. If $\Omega'$ is obtained
from $\Omega$ by moving the particle at site $i$ to $i+\delta$, then
\begin{equation}
k(\Omega,\Omega')
=
\nu\,n_i(1-n_{i+\delta})
\exp\left[
\frac{\beta f a}{2}\tau_i\delta
\right].
\label{eq:interacting-hop-rate}
\end{equation}
If $\Omega'$ comes from $\Omega$ by flipping the orientation at site
$i$, then
\begin{equation}
k(\Omega,\Omega')
=
\alpha n_i .
\label{eq:interacting-tumble-rate}
\end{equation}
All other transition rates are zero.
The master equation governing the time-evolution of the probability $\mathcal P(\Omega,t)$ for the system to be in the configuration $\Omega$ at time $t$, is given by $\dot {\mathcal{P}}(\Omega,t)=\mathcal{L}^\dagger \mathcal P(\Omega,t)$, where
\begin{align}
    \mathcal{L}^\dagger \mathcal P(\Omega,t)&=\sum_{\Omega'\neq \Omega}
\left[
k(\Omega',\Omega)\mathcal P(\Omega',t)
-
k(\Omega,\Omega')\mathcal P(\Omega,t)
\right],
\label{eq:interactingGen}
\end{align}
where the sum is over all allowed transitions, as defined above. We denote the stationary distribution of the interacting system at temperature $T$ by $P_T(\Omega)$, $\mathcal{L}^\dagger P_T=0$.

\begin{figure}
    \centering
    \includegraphics[width=0.45\linewidth]{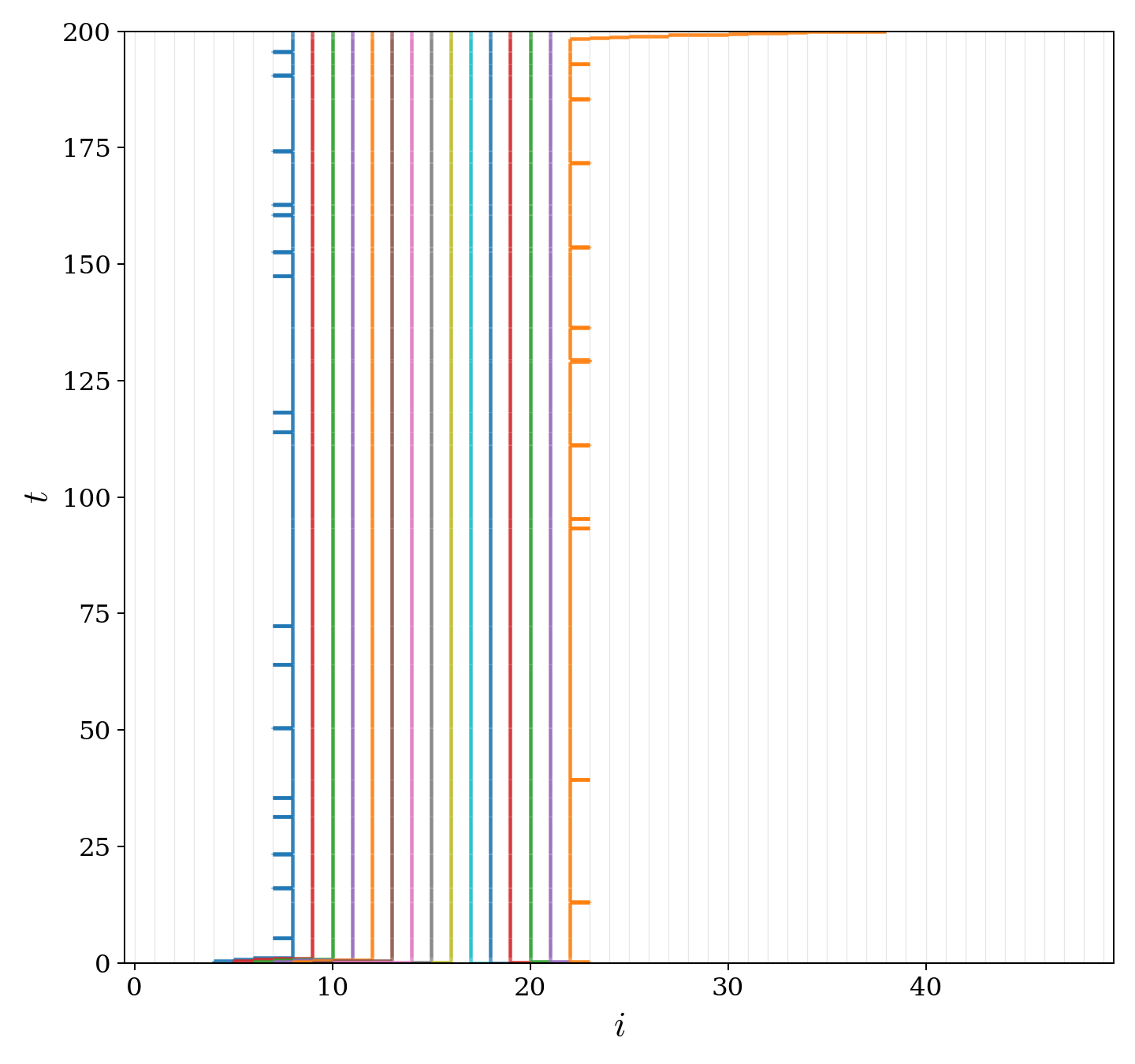}
    \includegraphics[width=0.45\linewidth]{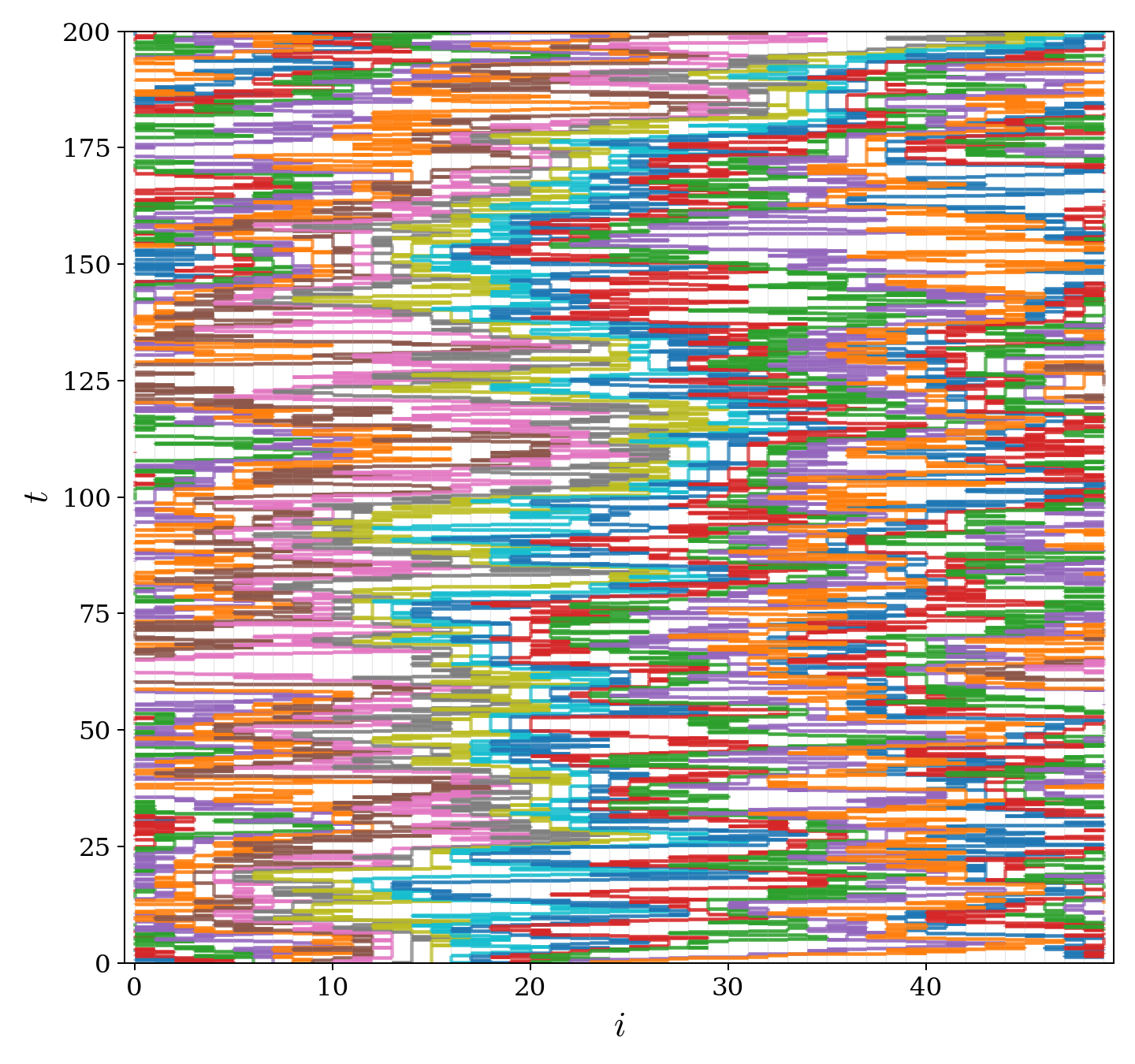}
\caption{Typical trajectories of $N_p=15$ interacting RTPs on a periodic
lattice with $L=50$ obtained from simulation of the dynamics.
Left: Jammed configuration for $\alpha=0.01$. Right: Homogeneous configuration for $\alpha=4$. Here we take $T=1,f=5$ for both plots.}
    \label{fig:trajinteracting}
\end{figure}

Run-and-tumble particles on one-dimensional lattices with exclusion have been studied before in the literature. It has been shown that they form dense jammed structures at high persistence~\cite{soto2014run,slowman2016jamming}. This can be understood in the following way. At small tumbling rate, two oppositely oriented particles can face each other and remain blocked for a long time, until one of them tumbles. Other particles can then accumulate behind them, forming a compact cluster. At large tumbling rate, orientations decorrelate rapidly, blocked configurations are short-lived, and the stationary state becomes much more homogeneous. 
This behavior is illustrated in the left panel of Fig.~\ref{fig:trajinteracting}. For small $\alpha$, the space-time trajectories show long-lived blocked regions. For large $\alpha$, these persistent runs are interrupted very frequently. The clusters then dissolve quickly, and the trajectories explore the ring more uniformly, as seen in the right panel of Fig.~\ref{fig:trajinteracting}.

\begin{figure}
    \centering
    \includegraphics[width=0.485\linewidth]{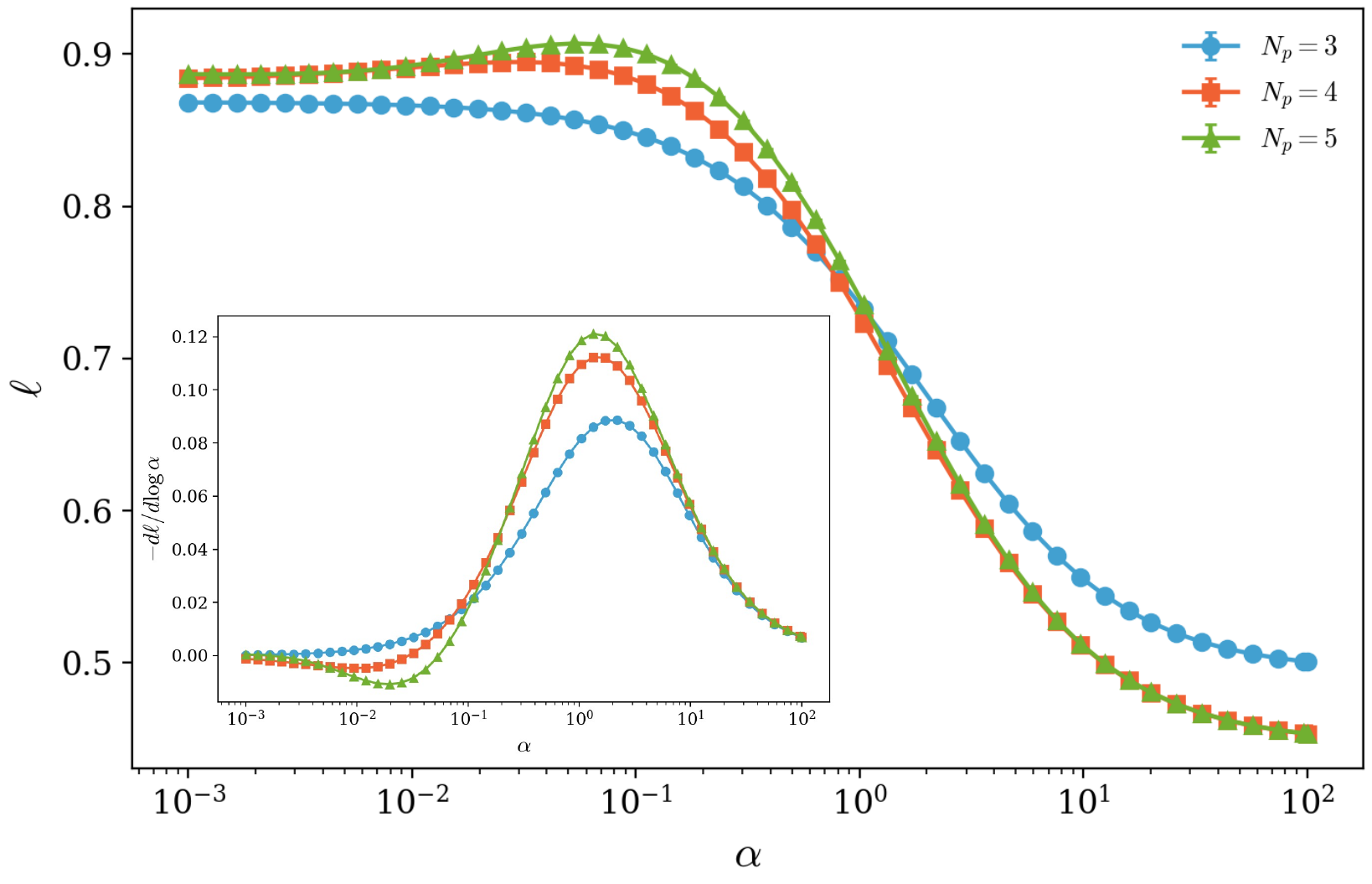}
     \includegraphics[width=0.485\linewidth]{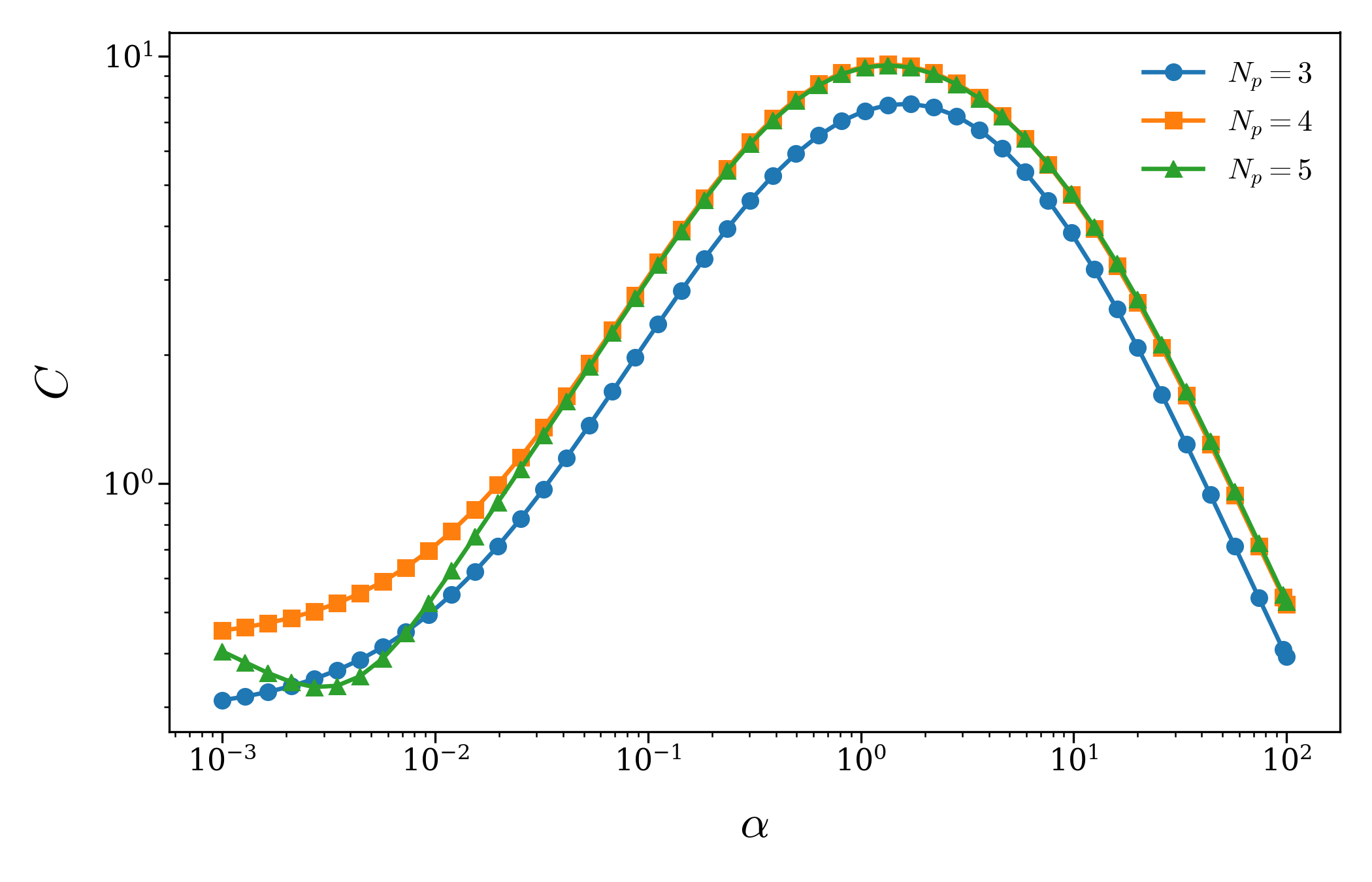}
    \caption{Left: Stationary mean block size $\ell$ as a function of
$\alpha$. The inset shows the corresponding $-d\ell/d\log\alpha$.  Right: heat capacity for the same parameters, namely,  $L=9$, $T=1$, $f=5$, and $a=1$.}
    \label{fig:clustersize-spheat}
\end{figure}

To quantify jamming, we measure the typical size of an occupied block. An
occupied block is a stretch of consecutive occupied sites, bounded on both
sides by empty sites. If the configuration $\Omega$ contains $N_b(\Omega)$
such blocks, we define
\begin{equation}
\lambda(\Omega)
=
\frac{N_p}{N_b(\Omega)} .
\label{eq:block-size-config}
\end{equation}
This is the mean block size in the configuration, since the $N_p$ particles
are divided among $N_b(\Omega)$ occupied blocks. The stationary normalized mean block
size is
\begin{equation}
\ell
=\frac 1{N_p}
\sum_{\Omega}P_T(\Omega)\lambda(\Omega).
\label{eq:mean-block-size}
\end{equation}
The left panel of Fig.~\ref{fig:clustersize-spheat} shows that the mean block size $\ell$ as a function of the activity $\alpha$ for a lattice of size $L=9$. The general trend is that the mean block size decreases as $\alpha$ increases. This confirms the crossover from a jammed active state at high persistence (small $\alpha$) to a homogeneous state for small persistence (large $\alpha$). It is worth noting that, there small increase in cluster size for $N_p>L/2$. In this regime there are only a few empty sites, and rare tumbles can help rearrange these gaps and merge nearby occupied blocks, thus a small increase of $\alpha$ increases the cluster size.  Beyond this initial regime, more frequent tumbling breaks up the jammed structures and the
cluster size decreases. Though the microscopic mechanism is different, a recent study of a one-dimensional active exclusion process also found that persistence affects jamming non-monotonically~\cite{jain2026jamming}.

\subsection{Excess heat}
For a succcessful spatial jump of a particle by $\delta=\pm 1$, with orientation $\sigma$, the heat released to the environment is $fa\sigma \delta$, while the tumblings carry no heat. The excess heat and nonequilibirum heat capacity then follow the same construction (Eqs.~\eqref{eq:dotq1}-\eqref{eq:spheatdef}) as in the single particle case with the single particle state $X$ replaced by $\Omega$, and the generator Eq.~\eqref{eq:interactingGen}.Similar to the single particle case, the centered instantaneous heat flux has a useful representation [see App.~\ref{app:1}] in terms of $N_b$ and,
\begin{align}
    B_\text{edge}(\Omega)=\sum_i \tau_i n_i (n_{i+1}- n_{i-1})
\end{align}
as,
\begin{align}
    h_T(\Omega)=fa(r-l)\delta N_b(\Omega)-\frac{fa(r+l)}{2}\delta B_\text{edge}(\Omega).
    \label{instheatLmulti}
\end{align}
Particles in the interior of a block have $n_{i+1}=n_{i-1}=1$, and do not contribute. Similarly, isolated particles $n_{i+1}=n_{i-1}=0$ also do not contribute. For blocks whose edge particles preferntially point inwards have $B_\text{edge}>0$, while for those whose edge particles point outwards have $B_\text{edge}<0$. Thus the heat source entering the Poisson equation has two components, associated with the relaxation of the number of occupied blocks ($N_b$) and the orientational state of their edges ($B_\text{edge}$).

We find that the heat capacity has one distinct maxima at an intermediate value of
$\alpha$. At small $\alpha$, the particles remain trapped for long times in jammed
configurations. Although a temperature perturbation changes the hopping
rates, most attempted moves are blocked by exclusion, and the stationary
organization changes only weakly. The corresponding excess heat is
therefore small. As $\alpha$ increases, the ends of the jammed clusters reorient more frequently, and the stationary states become more sensitive to temperature, producing the distinct peak. At much larger values of  $\alpha$, the orientations decorrelate rapidly and the configurations
become homogeneous, weakening the temperature dependence again. 

For $N_p=5$, the heat capacity shows an additional minimum at small $\alpha$, in the same regime where the mean cluster size initally increases. This correspondence suggests that the temperature-induced
rearrangements which promote the formation of extended jammed structures
lower the calorimetric response. Both the initial enhancement
of clustering and its subsequent loss therefore leave signatures in the
calorimetric response.

\section{Dependence on the kinetic pre-factor}\label{sec:kineticfactor}
In the above discussion we chose the kinetic pre-factor $\nu$ [see Eq.~\eqref{eq:single-rate}] to be unity. Local detailed balance, constrains the time-antisymmetric part of the rates, associated with entropy flux, but leaves freedom in their time-symmetric part $\nu$, which multiplies the forward and backward transition rates equally. This kinetic contribution, often referred to as dynamical activity or frenesy, can influence stationary occupations and nonequilibrium response even when the thermodynamic driving is unchanged~\cite{maes2021local,frenesy,maesresponse}. For a general temperature dependent $\nu(T)$, the master equation is modified in the following way for the single particle case,
\begin{align}
\dot p_i(t) &=
\nu(T)\left[R_{i-1}^{+}p_{i-1}(t)
+
L_{i+1}^{+}p_{i+1}(t)
-
(R_i^+ + L_i^+)p_i(t)
+
\alpha'(m_i(t)-p_i(t))\right],
\label{eq:p-master2}
\\
\dot m_i(t) &=
\nu(T)\left[R_{i-1}^{-}m_{i-1}(t)
+
L_{i+1}^{-}m_{i+1}(t)
-
(R_i^- + L_i^-)m_i(t)
+
\alpha'(p_i(t)-m_i(t))\right],
\label{eq:m-master2}
\end{align}
where $\alpha'=\alpha/\nu(T)$. Thus the corresponding stationary state can be obtained just by rescaling the tumbling rate by $\nu(T)$. The instantaneous heat flux in Eq.~\eqref{eq:dotq1} is also multiplied by $\nu(T)$. Since the generator and the heat flux acquire the same overall factor, this factor cancels from the Poisson equation. The quasipotential therefore also depends on the kinetic prefactor only through the scaled tumbling rate $\alpha'$. 

If $\nu$ is independent of temperature, the complete calorimetric curve is obtained by the same rescaling $\alpha\to\alpha'$. When $\nu(T)$ depends on temperature, however, a temperature perturbation also changes $\alpha'$. Indeed, 
\begin{equation}
\left.\frac{\partial P_T(X)}{\partial T}\right|_{\alpha}
=
\left.\frac{\partial P_T(X)}{\partial T}\right|_{\alpha'}
-
\frac{d\log\nu}{dT}
\frac{\partial P_T(X)}{\partial\log\alpha'}.
\label{eq:prob-temperature-prefactor}
\end{equation}
Thus the heat capacity gives \begin{align} 
C =&
\sum_X
\left.\frac{\partial P_T(X)}{\partial T}\right|_{\alpha'}
V_T(X)-
\frac{d\log\nu}{dT}
\sum_X
\frac{\partial P_T(X)}{\partial\log\alpha'}
V_T(X).
\label{eq:specific-heat-prefactor}
\end{align} 
The first term is the calorimetric response at fixed relative hopping and tumbling timescales. The second term appears because changing the temperature also changes the hopping rate relative to tumbling. It contains the effects of the kinetic pre-factor $\nu(T),$ and is controlled by the change of the stationary distribution with $\alpha'$, weighted by the quasipotential. This contribution has no fixed sign and can
change not only the magnitude, but also the shape and sign of the specific
heat.

\begin{figure}
    \centering
    \includegraphics[width=0.85\linewidth]{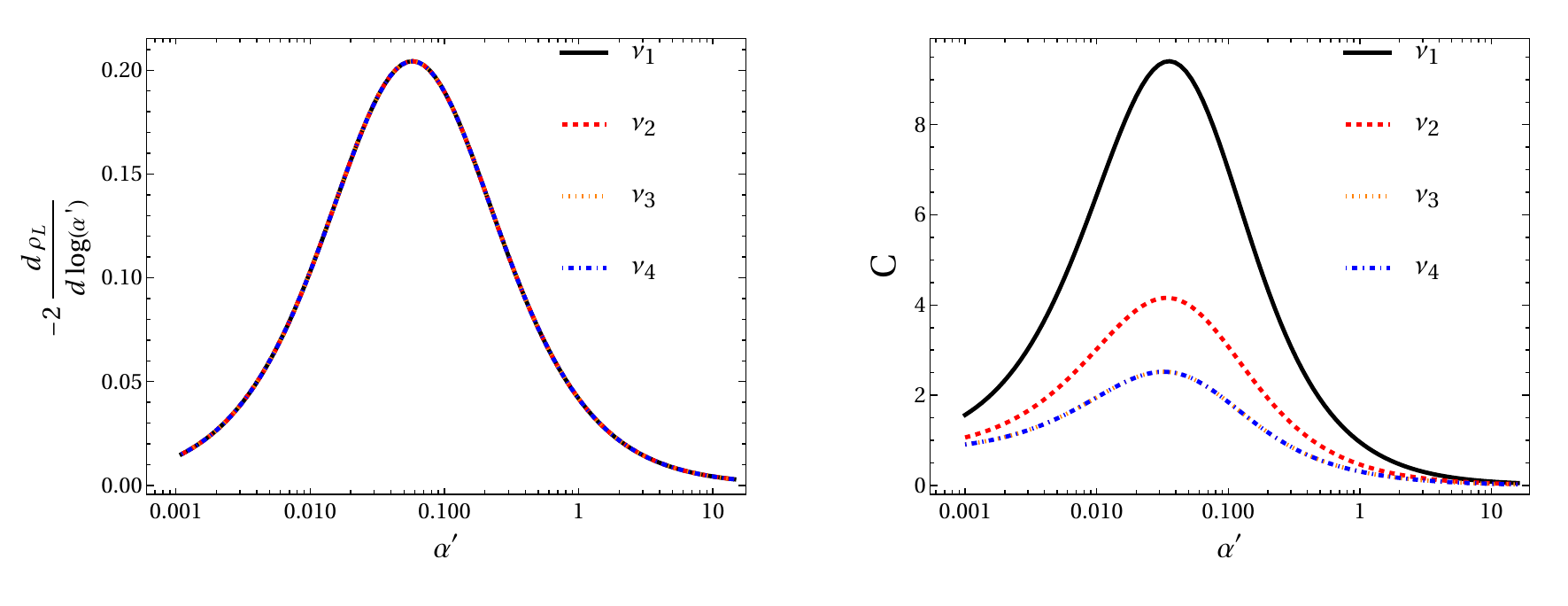}
    \caption{Effects of changing the kinetic prefactor on the stationary configuration (left), and heat capacity (right) of the non-interacting active particle on a bounded lattice $[-L,L]$. Both plots use use $L=20$, $N_p=1$, $T=1$, $f=2$, and $a=\gamma=1$. }
    \label{fig:kineticsp}
\end{figure}

We illustrate the effects of the kinetic prefactor, using three additional choices for $\nu(T)$, namely, 
\begin{align}
\nu_1(T)=1,\quad    \nu_2(T)=\frac{1}{\cosh( fa/(2T))},\quad\nu_3(T)=e^{-fa/(2T)},\quad \nu_4(T)=\gamma T/a^2.
\end{align}
The first choice is the same as in the previous section, while the last one is such that the continuum Langevin equation reduces to the usual RTP with thermal noise. We find that the strucural crossover is not affected when plotted against the scaled tumbling rate [see left panel of Fig.~\ref{fig:kineticsp}], consistent with our predictions above. The heat capacity, on the other hand, has different amplitudes and shapes due to the additional term in Eq.~\eqref{eq:specific-heat-prefactor}, as shown in Fig.~\ref{fig:kineticsp} (right panel). However, there is always a clear feature  corresponding to the jammed to unjammed transition, but the precise position and magnitude are not universal. 

\begin{figure}
    \centering
    \includegraphics[width=0.85\linewidth]{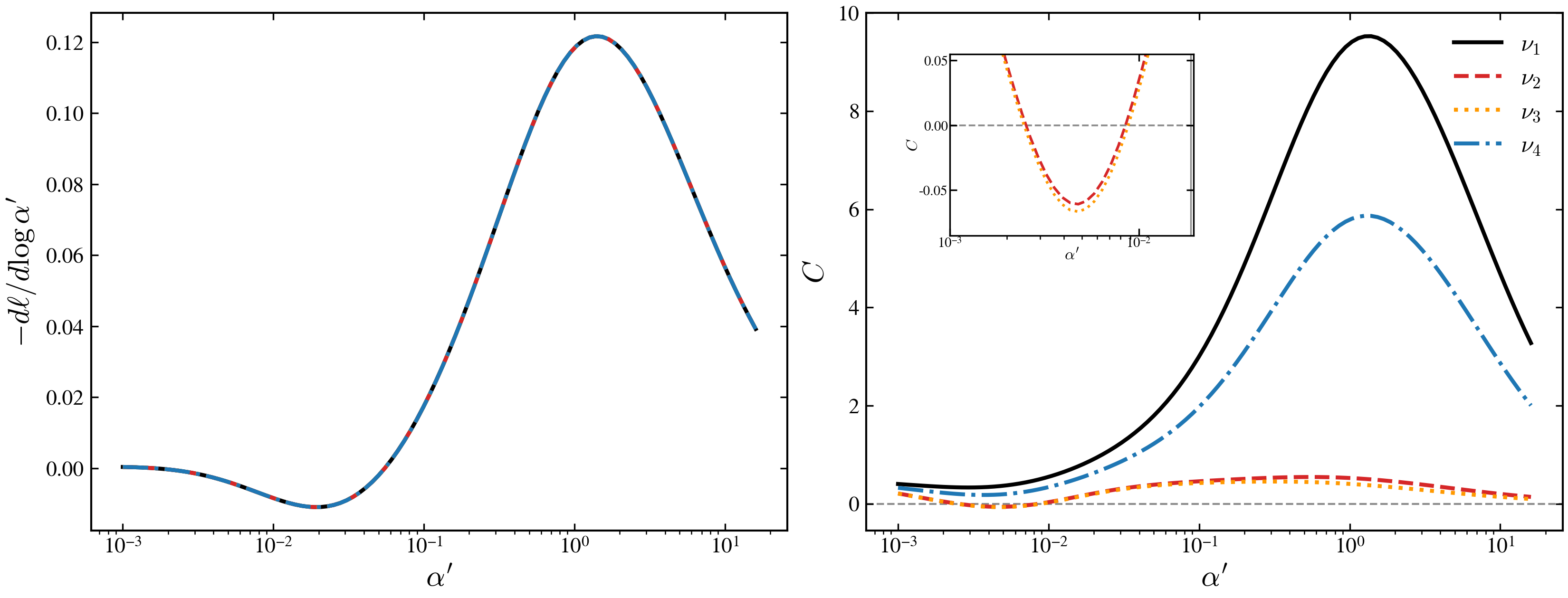}
    \caption{Effects of changing the kinetic prefactor on the stationary density (left), and heat capacity (right) of $N_p=5$ active particles with exclusion on a periodic lattice $L=9$. Both plots use use $T=1$, $f=5$, and $a=1$. The inset on the right plot zooms near the extremum at small $\alpha'$ which becomes negative for $\nu_2$, and $\nu_3$.}
    \label{fig:kineticint}
\end{figure}

We next ask whether this separation between a trivially rescaled stationary
structure and a nontrivially modified calorimetric response persists in the
interacting system. The interacting system shows the same general behavior. The jamming properties are unchanged as functions of $\alpha'$, as shown in the left panel of Fig.~\ref{fig:kineticint}, whereas the heat capacity depends strongly on the choice of $\nu(T)$ [see right panel of Fig.~\ref{fig:kineticint}]. Different prefactors may produce peaks, dips, or even negative heat capacities [see inset of the right panel in Fig.~\ref{fig:kineticint}], but the strongest calorimetric variations remain associated with the reorganization of the jammed state.

\section{Experimental possibilities}
\label{s:accal}
The well-defined lattice dynamics enabled us to write the dynamics in terms of a Markov generator, which facilitated the computations in the previous sections. It is natural to ask whether the same quantity can be accessed in experimental systems, where such active to passive crossovers have been observed. AC calorimetry provides such an operational route. The method is well established for equilibrium systems, where a weak periodic heating or temperature modulation allows the heat capacity to be extracted from the
amplitude and phase of the thermal response. The same idea can be extended to nonequilibrium stationary systems, provided the continuously dissipated housekeeping heat is separated from the relaxational excess heat.

Following the nonequilibrium AC-calorimetry formulation of~\cite{calo}, we consider a weak periodic perturbation $g(t)=\varepsilon \cos(\omega t)$ around $T$,
where $\varepsilon/T\ll 1$. 

The response of the instantaneous mean heat flux  $\dot Q_t=\sum_X \mathcal{P}(X,t)\dot Q_{T(t)}(X)$ can be written as,
\begin{equation}
    \la\dot Q(t)\ra=\la\dot Q_T\ra+R_{\text{inst}}g(t)+\int_0^t dt' R(t-t')g(t')
\end{equation}
Here, $R_{\text{inst}}$ is the instantaneous response due to the explicit temperature dependence of the transition rates, while $R(t)$ describes the delayed response arising from the relaxation of the probability distribution. After the initial transient has decayed, substituting
$g(t)=\varepsilon\cos(\omega t)$ gives
\begin{align*}
\la\dot Q(t)\ra-\la\dot Q_T\ra
=
\varepsilon
\left[
R_{\text{inst}}
+
\int_0^\infty dt'\,R(t')\cos(\omega t')
\right]\cos(\omega t)
+
\varepsilon
\left[
\int_0^\infty dt'\,R(t')\sin(\omega t')
\right]\sin(\omega t).
\end{align*}
The first term is in phase with the imposed temperature modulation, while
the second is shifted by $\pi/2$. At low frequencies,
\begin{equation}
\la\dot Q(t)\ra
=
\la\dot Q_T\ra
+
\varepsilon
\left[
B(T)\cos(\omega t)
+
\omega C^{\text{AC}}(T)\sin(\omega t)
+
O(\omega^2)
\right],
\label{eq:ac-low-frequency-response}
\end{equation}
where
\begin{equation}
B(T)
=
R_{\text{inst}}
+
\int_0^\infty ds\,R(s)
=
\frac{d}{dT}\la\dot Q_T\ra.
\end{equation}
Using $\int_0^\infty ds\,R(s)\sin(\omega s)
=
\omega\int_0^\infty ds\,sR(s)+O(\omega^3),$
the out-of-phase coefficient is,
\begin{equation}
C^{\text{AC}}(T)
=
\int_0^\infty ds\,sR(s),
\label{eq:ac-response-coefficients}
\end{equation}
which equals the nonequilibrium heat capacity defined from the excess heat
in Eq.~\eqref{eq:spheatdef}. 
Thus, the quasistatic heat capacity appears as the leading low-frequency out-of-phase response of the heat flux. This is the usual principle of AC calorimetry, extended here to a nonequilibrium stationary state with a nonzero stationary heat dissipation.
\begin{figure}
    \centering
    \includegraphics[width=0.48\linewidth]{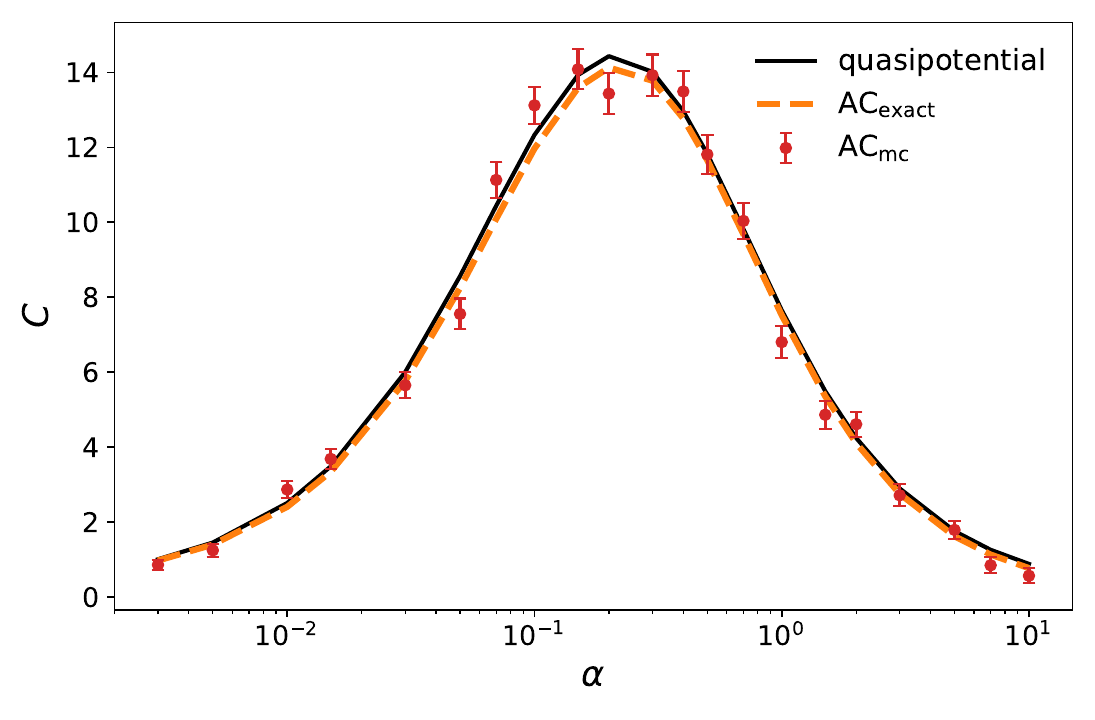}
\includegraphics[width=0.48\linewidth]{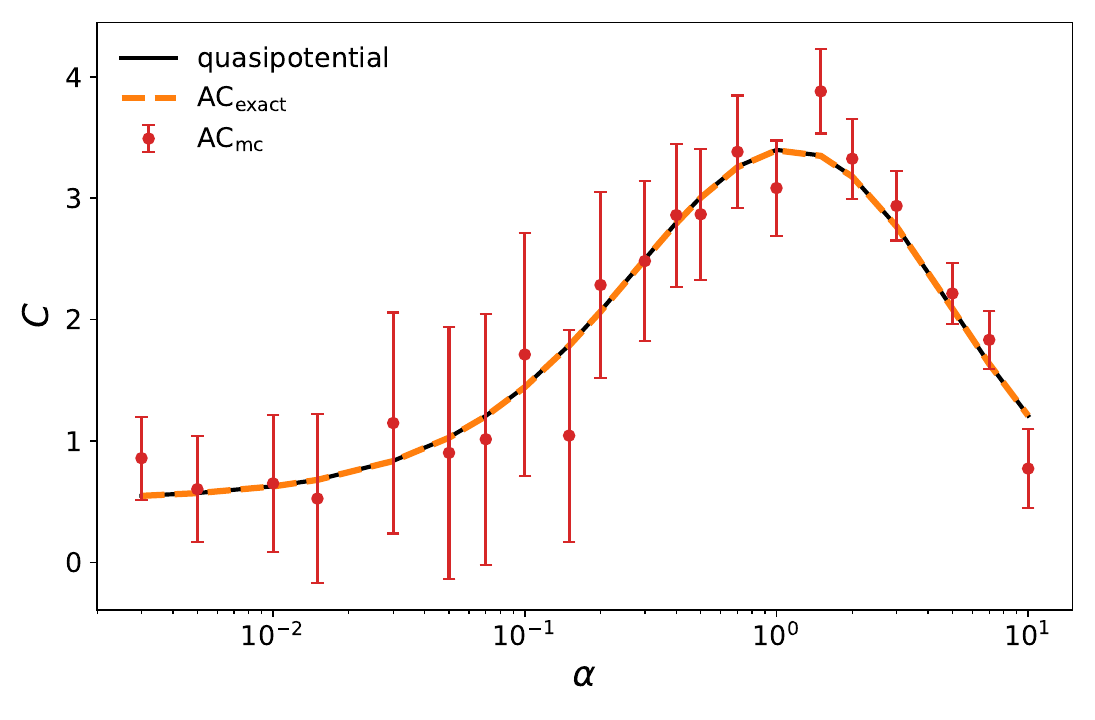}
    \caption{Comparison of the heat capacity computed using Monte-Carlo simulations for the AC calorimetry protocol [in symbols], with the same obtained from the quasipotential calculation [in solid lines], and exact AC calorimetry results (see Eq.~\eqref{eq:ac-exact}) [in dashed lines]. Here for all the curves $\nu=1,\,T=1,\,a=1$, and for the last AC curves $\omega=0.25$. Left panel: $L=20$, $\lambda=0$; right panel: same parameters as Fig.~\ref{fig:clustersize-spheat}.}
    \label{f:accalorie}
\end{figure}

We also test this for the models with $\nu=1$, and they are consistent with the heat capacities obtained via the Poisson equation, for sufficiently small driving frequencies. The results for the single particle and interacting particles are shown in Fig.~\ref{f:accalorie} left and right panels, respectively. This demostrates that it might also be possible to obtain these calorimetric signatures in experiments in the spirit of~\cite{bottcher1997microcalorimetry,vehusheia2023microcalorimetry}.

\section{Conclusion}\label{s:concl}
We have shown that structural transitions in active-particle systems leave pronounced signatures in the nonequilibrium heat capacity. For a confined single particle, the loss of boundary accumulation is accompanied by a clear calorimetric feature, while for interacting particles the reorganization of jammed clusters produces a corresponding nonmonotonic heat response. The precise shape and sign of this response depend on the kinetic details of the dynamics, but its strongest variations remain associated with the regime in which the stationary structure changes most rapidly. Nonequilibrium heat capacity can therefore serve as a thermodynamic probe of structural reorganization in active systems. 

Several extensions are natural. The interacting calculations presented here
are restricted to small system sizes by the exponential growth of the configuration
space. More efficient numerical approaches, such as tensor-network methods~\cite{strand2022computing,strand2022using}, could
allow the crossover and its calorimetric signature to be studied in larger system sizes, and possibly in the thermodynamic limit. This could lead to the identification of critical exponents, and universality classes associated with these transitions.
Extending the analysis to two and higher dimensions would also make it possible to connect the excess-heat response more directly with clustering and
motility-induced phase separation. Another interesting avenue would be to see if our predictions can be detected in experiments with AC calorimetry.

\section*{Acknowledgments}
The author thanks Christian Maes for useful discussions and comments, and the European Union’s Horizon 2024 research and innovation programme under the Marie Sklodowska–Curie (HORIZON-TMA-MSCA-PF-EF) grant agreement No. 101205210 for funding. 
\appendix

\section{Simplified forms of instanataneous heat flux}\label{app:1}
Here we derive the representations of the centered instantaneous heat flux
$h_T$ used in the main text. Throughout this appendix we set $U_i=0$ and
$\nu=1$, and thus have two rates $ r=e^{\beta fa/2},$ and $l=e^{-\beta fa/2}$.
\subsection*{Single particle}
For the single-particle system, the instantaneous heat flux conditioned on
the state $X=(i,\sigma)$ is given by Eq.~\eqref{eq:dotq1}, $\dot Q_T(X) = \sum_{\delta=\pm1}   k(i,\sigma\to i+\delta,\sigma)\, fa\sigma\delta ,$ where transitions outside $[-L,L] $ are omitted. In the bulk 
\begin{equation}
\dot Q_T(X)=fa(r-l),    \qquad |i|<L.
\end{equation} 
For the boundaries, we introduce $B=B_\text{out}+B_\text{in}$, and $M=B_\text{out}-B_\text{in}$ with  
the indicator functions $B_\text{out}=\delta_{i,L}\delta_{\sigma,+}+\delta_{i,-L}\delta_{\sigma,-}$, and $B_\text{in}=\delta_{i,L}\delta_{\sigma,-}+\delta_{i,-L}\delta_{\sigma,+}$. $B(X)$ tells us whether the particle is at a boundary, while $M(X)$ outward from inward orientation. The instantaneous heat flux can
then be written for all states as
\begin{align}
    \dot Q_T(X)
    &=
    fa(r-l)
    -fa\,r B_{\rm out}(X)
    +fa\,l B_{\rm in}(X)\\
    &= fa(r-l)
    -
    \frac{fa}{2}
    \left[
        (r-l)B(X)
        +(r+l)M(X)
    \right].
    \label{app:single-dotQ}
\end{align}
Upon subtracting the stationary average, we get the expression Eq.~\eqref{inst:excess-single}.

\subsection{Interacting particles}
The instantaneous heat flux follows 
\begin{equation}
\dot Q_T(\Omega)
=
fa
\sum_{i=1}^{L}\sum_{\delta=\pm1}
n_i(1-n_{i+\delta})
\,\tau_i\delta\,
e^{\frac{\beta fa}{2}\tau_i\delta}.
\label{app:multi-dotQ-start}
\end{equation}
For $\tau_i\delta=\pm1$, one has
\begin{align}
\dot Q_T(\Omega)
={}&
\frac{fa(r-l)}{2}
\sum_{i,\delta}
n_i(1-n_{i+\delta})
+
\frac{fa(r+l)}{2}
\sum_{i,\delta}
n_i(1-n_{i+\delta})\tau_i\delta .
\label{app:multi-split}
\end{align}
The first sum counts particle-vacant space interface, every occupied block has two such interfaces and $\sum_{i,\delta}
    n_i(1-n_{i+\delta})
    =
    2N_b(\Omega)$
The second sum is $\sum_{i,\delta}
n_i(1-n_{i+\delta})\tau_i\delta
=
-\sum_i
\tau_i n_i
(n_{i+1}-n_{i-1})$. Using these we get,
\begin{equation}
    \dot Q_T(\Omega)
    =
    fa(r-l)N_b(\Omega)
    -
    \frac{fa(r+l)}{2}B_{\rm edge}(\Omega),
    \label{app:multi-dotQ}
\end{equation}
with $B_{\rm edge}(\Omega)
    =
    \sum_i
    \tau_i n_i
    (n_{i+1}-n_{i-1}).$
Particles in the interior of a block have
$n_{i+1}=n_{i-1}=1$ and do not contribute to $B_{\rm edge}$.
Similarly, isolated particles have $n_{i+1}=n_{i-1}=0$ and also do not
contribute. Thus $B_{\rm edge}$ only probes the orientation of particles at
the edges of occupied blocks. Finally subtracting the stationary average, we get the expression quoted in the main text.

\section{Details of numerics}
\label{app:numerics}
\subsection{Matrix computations}
For both the single- and interacting-particle systems, we construct the
forward generator $\mathcal{L}^{\dagger}$ in the corresponding finite state
space and store it as a sparse matrix, since each state is connected by the
dynamics to only a small number of other states. The stationary distribution is then
obtained from $\mathcal{L}^{\dagger}P_T=0$ with the normalization $\sum_X P_T(X)=1.$
The quasipotential is obtained by solving $    \mathcal{L}V_T=-h_T$, with
  $  \sum_XP_T(X)V_T(X)=0.$. The temperature derivative entering the heat capacity is evaluated using a centered finite difference,
\begin{equation}
    \frac{\partial P_T(X)}{\partial T}
    \simeq
    \frac{P_{T+\Delta T}(X)-P_{T-\Delta T}(X)}
         {2\Delta T},
\end{equation}
with $\Delta T=0.01$. The heat capacity is then evaluated from Eq.~\eqref{eq:spheatdef}.

\subsection{Exact finite-frequency AC response}
For the periodic temperature modulation discussed in the main text,  expanding the probability distribution is written to linear order in $\varepsilon$, $P(t)=P_T+\varepsilon\,\mathrm{Re}
\left[\tilde P\,e^{i\omega t}\right].$ The corresponding equation in Fourier space is
\begin{equation}
\left(i\omega-\mathcal{L}^{\dagger}\right)\tilde P
=
\left(\partial_T\mathcal{L}^{\dagger}\right)P_T,
\qquad
\sum_X\tilde P(X)=0,
\label{eq:ac-frequency}
\end{equation}
where $\partial_T\mathcal{L}^{\dagger}$ is evaluated by a centered finite
difference. Using the out-of-phase response defined in
Eqs.~\eqref{eq:ac-low-frequency-response} and
\eqref{eq:ac-response-coefficients}, the finite-frequency heat capacity is
obtained as
\begin{equation}
C^{\rm AC}(\omega)
=
-\frac{1}{\omega}
\sum_X
\dot Q_T(X)\,
\operatorname{Im}\tilde P(X).
\label{eq:ac-exact}
\end{equation}

\subsection{Monte Carlo results}
For the Monte Carlo results in Fig.~\ref{f:accalorie}, we simulate the
periodically driven Markov jump process with the transition rates evaluated
at the instantaneous temperature $T(t)$. The conditional heat current is
given by Eq.~\eqref{eq:dotq1}. The dynamics is generated using continuous-time
kinetic Monte Carlo with a thinning procedure~\cite{lewis1979simulation} to
account for the explicit time dependence of the transition rates. After the
initial transient, the out-of-phase component is measured over an integer
number of driving periods,
\begin{equation}
    J_{\sin}
    =
    \frac{2}{t_{\rm meas}}
    \int_0^{t_{\rm meas}}
    dt\,j_Q(t)\sin(\omega t),
\end{equation}
and the AC heat capacity follows from Eq.~\eqref{eq:ac-low-frequency-response}
as
\begin{equation}
    C^{\rm AC}=\frac{J_{\sin}}{\varepsilon\omega}.
\end{equation}

A parent trajectory is first relaxed for $64$ driving periods. Configurations
are then sampled every four complete periods, so that all branches start at
the same phase of the temperature modulation. For the single-particle results,
we use $1024$ independent branches, each measured for $1024$ periods. For the
interacting system, we use $8192$ branches, each measured for $512$ periods.
The error bars are obtained from the standard error over the independent
branches.

\bibliographystyle{apsrev4-2}
\bibliography{ref0}
\end{document}